\documentclass[aps,prl,twocolumn,showpac, reprint]{revtex4-1}

\usepackage{amssymb}
\usepackage{graphicx}
\usepackage{amsmath}
\usepackage{subfigure}
\usepackage{color}
\usepackage[export]{adjustbox}

\usepackage[normalem]{ulem}

\newcommand{\figpath}{}

\begin{document}

\title{Discrete self-similarity in interfacial hydrodynamics
and the formation of iterated structures}
\author{Michael C. Dallaston$^1$, Marco A. Fontelos$^2$, Dmitri Tseluiko$^3$, Serafim Kalliadasis$^4$}

\affiliation{$^1$School of Computing, Electronics and Mathematics, and Flow Measurement and Fluid Mechanics Research Center,
Coventry University, \break Coventry CV1 5FB, UK\\
$^2$Instituto de Ciencias Matem\'{a}ticas, C/Nicol\'{a}s Cabrera, Madrid, 28049, Spain\\
$^3$Department of Mathematical Sciences, Loughborough University, \break Loughborough LE11 3TU, UK\\
$^4$Department of Chemical Engineering, Imperial College London, \break London SW7 2AZ, UK}

\begin{abstract}
    The formation of iterated structures, such as satellite and
    sub-satellite drops, filaments and bubbles, is a common feature in
    interfacial hydrodynamics. Here we undertake a computational and
    theoretical study of their origin in the case of thin films of viscous
    fluids that are destabilized by long-range molecular or other forces.
    We demonstrate that iterated structures appear as a consequence of
    discrete self-similarity, where certain patterns repeat themselves,
    subject to rescaling, periodically in a logarithmic time scale. The
    result is an infinite sequence of ridges and filaments with similarity
    properties. The character of these discretely self-similar solutions as
    the result of a Hopf bifurcation from ordinarily self-similar solutions
    is also described.
\end{abstract}

\maketitle

% NOTE: A full address must be provided: department, university/institution, town/city, zipcode/postcode, country.

Free-surface flows can produce a great diversity of patterns such as
filaments, drops, bubbles, pearls, etc.~\cite{EV}.
Amongst them, probably the most intriguing and elusive to
analyze have been the so called ``iterated patterns'', i.e. ``patterns within
patterns" where the same structure repeats itself at different time and
length scales. Such structures appear in a wide variety of physical,
biological and technological settings: from natural phenomena with fractal
features to elasticity and composite
materials~\cite{General}. In the context of interfacial hydrodynamics, in
particular, examples of iterated structures are the formation of several
generations of satellite drops in capillary breakup~\cite{TSO}, the cascade
of structures produced in viscous jets \cite{SBN12} and the
iterated stretching of viscoelastic filaments \cite{CDKOM}. In this letter,
we present for the first time a scenario
%{in which stability analysis and careful numerical computation reveal}
revealing how such structures may appear via a bifurcation
%, as some structural parameter changes,
from self-similar solutions to discretely self-similar ones where scale
invariance occurs only at discrete times, resulting in the infinite
repetition of some pattern at a discrete sequence of time and length scales.
 Discrete self-similarity has proven to be present at some
instances of gravitational collapse~\cite{Choptuik} and has also been
proposed as a mechanism for the development of turbulence and formation of
singularities in Euler's equation through chaotic
self-similarity~\cite{Pumir-Siggia}. Our study reveals the mechanism for
discrete self-similarity and ensuing complexity on all scales via a model
system consisting of a reduced-order hydrodynamic evolution equation.

The physical situation we consider is the rupture of thin films driven by a
destabilizing effect.  Liquid films are ubiquitous in a wide spectrum of
natural phe\-no\-me\-na and technological applications \cite{CM}. One
well-studied effect is that of long-range intermolecular or van der Waals
forces; when the film is sufficiently thin, these forces may cause the film
to destabilize and eventually rupture and dewet the substrate. In the
long-wave approximation (appropriate for slow flows with strong surface
tension), the problem may be formulated in terms of an evolution equation for
the film profile $h(\mathbf{x},t)$ in the form $ h_{t}=-{\boldsymbol
\nabla}\cdot\mathbf{q}, $
 where $\mathbf{q}=-({h^{3}}/{3\mu }){\boldsymbol \nabla} p$ is the flow rate, with $\mu $
being the liquids's viscosity and $p=-\sigma \nabla ^{2}h-\Pi (h)$ being the pressure.
%
%\begin{equation*}
%-p=\sigma \nabla ^{2}h+\Pi (h).
%\end{equation*}%
The first $p$ component is the Laplace pressure (surface tension times linearized curvature) and the second is the disjoining pressure taken
to be
%of the form
$\Pi (h)=-{A}/{h^{n}}.$ %
%\begin{equation*}
%\Pi (h)=-\frac{A}{h^{n}}.
%\end{equation*}%
In one spatial dimension, and after a suitable nondimensionalization,
%the model for the evolution of a thin liquid film under the action of molecular forces then reads%
the evolution equation for $h$ reads
\begin{equation}
h_{t}+\biggl[ h^{3}\left( h_{xx}-\frac{1}{nh^{n}}\right) _{x}\biggr] _{x}=0.
\label{vdw}
\end{equation}
In the context of rupture by van der Waals forces, $A$ (strictly, $6\pi A$)
is the Hamaker constant, while $n$ is almost always taken to be
$3$ \,\cite{ZL,WB12}.
%{\color{red}\bf (DT: I've removed reference to \cite{YSKYPKDNGT} here, as it is anyway discussed in detail below.)}

At $n=3$, a remarkable property of solutions of (\ref{vdw}) is the
development of self-similar film rupture ($h\rightarrow 0$ at a single point) in finite
time\,\cite{ZL,WB12}. There is, however, good reason to examine different
values of $n$. {Yatshishin \emph{et
al.}~\cite{YSKYPKDNGT} %in Ref.~\cite{YSKYPKDNGT}
show that the disjoining pressure with $n = 3$ is an asymptote to DFT as the
distance of the chemical potential from saturation vanishes, assuming a
Lennard--Jones potential for pair\-wise molecular interactions and neglecting
screening effects. Thus, the usual form with $n=3$ is only approached for
thick films. For thin films, there is a deviation from the $n=3$ behavior,
and, dependent on the system, different exponents might be possible.
Furthermore, the behavior on $h$ might be non-local, a consequence of the
non-local character of the long-range intermolecular interactions (see
also~\cite{BEIMR,Parry_etal}). However a widely-adopted algebraic dependence
offers ease of access to the corresponding equations facilitating their
analytical-numerical scrutiny. At the same time, a great deal of experimental
study has shown that the functional form of $\Pi(h)$ is highly dependent on
the nature of the dominant intermolecular force, which is influenced by the
substrate and liquid properties. For example, \cite{DC} found that the
effects of screening can lead to a potential better modeled with $n=4$ in
hydrocarbon-metal experiments with $h>40$\,nm, while the contribution due to
hydrogen bonding in water-silica-glass experiments for very thin ($<30$\,nm)
films was better approximated by $n=1$ {(Pashley~\cite{PD}). For water on
quartz, $\Pi(h)$ is estimated to also have $n=1$ for $h < 80$\,nm and $n=2$
for $h>120$\,nm~\cite{PD}. For $80\,\text{nm}<h<120\,\text{nm}$, it would
then be appropriate to take $1<n<2$. In any case, given a liquid and a
substrate, we
can approximate, when appropriate, $\Pi(h)$ for relatively thin films with a power law by fitting a %an appropriate
value of~$n$. We emphasize that assuming a certain form for the disjoining
pressure and fitting appropriate values of its parameters is common in the
literature (e.g.~\cite{HTA}). There has also been recent interest in
generalizing the standard Lennard--Jones potential with attractive exponent
$\lambda_a = 6$ to other exponents, leading to the so-called Mie potential
\cite{RAMG}; the corresponding disjoining pressure has exponent $n =
\lambda_a - 3$~\cite{I}. A good summary of different contributions to the
disjoining pressure may be found in \cite{TDS12}. The dynamics of rupture
under these different values has not previously been examined.

\begin{figure}[t]%[tbph]
\centering
\includegraphics[width=0.415\textwidth]{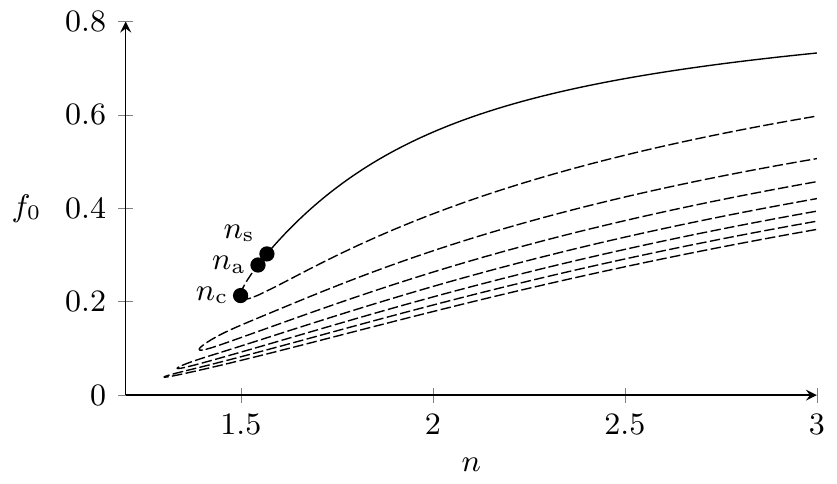}
\vspace{-0.4cm}
\caption{{}Bifurcation diagram for self-similar solutions $f(\xi)$ satisfying (\ref{sseq}), labelled by $f_{0}=f(0) $. For sufficiently large $n$, there are infinitely many solutions. At $n=n_{\mathrm c} \simeq 1.49915$ there is a first turning point where the first and
second branches of solutions merge. Successive turning points exist at $1.39141$%
, $1.33405$, $1.29993$, \ldots.  The primary branch is stable for most of its domain; however, it becomes unstable due to a Hopf bifurcation pair $n_s \approx 1.567$, $n_a \approx 1.545$ (see Fig.~\ref{fig5a}).}
\label{fig1}
\vspace{-0.5cm}
\end{figure}

As well as intermolecular forces, (\ref{vdw}) models
other thin-film phenomena at different scales, such as destabilization
due to thermocapillarity~\cite{BTARSKRJ} and density contrast
(Rayleigh--Taylor instability)~\cite{YH}; in such cases we may define an
equivalent ``disjoining pressure'' behaving as $\ln(h)$ for
the thermocapillary effect (essentially $n=0$), or as $h$ for the
Rayleigh--Taylor instability ($n=-1$).  Instead of self-similar rupture,
these two examples exhibit cascades of satellite droplets, similar to those
discussed above, so it is of great interest to understand how the two
behaviors are connected through variation in $n$.

Assuming that rupture occurs at a single
point $x_{0}$ at time $t_{0}$, it is natural to seek solutions in
a coordinate system that focuses on the point and time of rupture:
\vspace{-0.2cm}
\begin{equation}
\!h(x,t)\!=\!(t_{0}\!-\!t)^{\alpha }f( \xi, \tau),\ \xi\! =\! \frac{x\!-\!x_{0}}{(t_{0}\!-\!t)^{\beta
}}, \ \tau\! =\! -\ln(t_0\!-\!t),\!\!\!\!\! \label{ss}
\vspace{-0.1cm}
\end{equation}%
%\begin{equation}
%\begin{aligned}
%&h(x,t)=(t_{0}-t)^{\alpha }f\left( \xi, \tau\right),\\   &\xi = \frac{x-x_{0}}{(t_{0}-t)^{\beta
%}}, \quad \tau = -\ln(t_0-t), \label{ss}
%\end{aligned}
%\end{equation}%
where, from simple dimensional arguments based on (\ref{vdw}), one finds
$\alpha ={1}/{(2n-1)},\ \beta ={(n+1)}/{(4n-2)}$.
%\begin{equation}
%\alpha =\frac{1}{2n-1},\ \beta =\frac{n+1}{4n-2}.  \label{ab}
%\end{equation}%
For a rupture solution to
exist, we must assume $n>{1}/{2}$, so that $\alpha,\,\beta>0$. The scaled profile $f(\xi,\tau)$
then satisfies%
\vspace{-0.2cm}
\begin{equation}
\label{sseq}
f_\tau = \alpha f - \beta \xi f_{\xi } - \biggl[ f^{3}\left( f_{\xi \xi
}-\frac{1}{nf^{n}}\right) _{\xi }\biggr] _{\xi },
\vspace{-0.1cm}
\end{equation}%
subject to the condition that the interfacial velocity $h_t$ remains finite
at a finite distance from $x_{0}$. As $t\rightarrow t_{0}$, one has $%
\xi \rightarrow \infty $ and, in order to cancel out singular dependence on $%
t_{0}-t$, we must impose%
\vspace{-0.2cm}
\begin{equation}
f_\tau \sim \alpha f - \beta \xi f_\xi, \qquad |\xi|\rightarrow\infty.  \label{bcvdw}
\vspace{-0.1cm}
\end{equation}
Steady states of (\ref{sseq}), (\ref{bcvdw}) represent self-similar solutions of~(\ref{vdw}).  Including $\tau$ dependence, allows us to examine the stability and dynamics in the vicinity of these solutions.

\begin{figure}[tbp]
\includegraphics[width=0.45\textwidth]{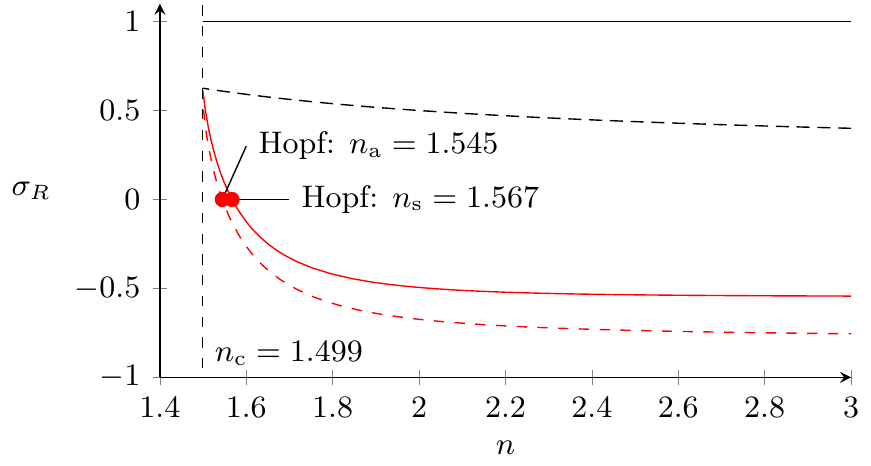}
\vspace{-0.4cm}
\caption{The real part $\sigma_R$ of eigenvalues governing the stabilty of the primary solution branch $f_1(\xi)$ as $n$ varies.  Symmetric and antisymmetric modes of perturbation are shown as solid and dashed lines, respectively.  The upper two eigenvalues are the trivial eigenvalues ($\sigma_1=1$, $\sigma_2=\beta$).  The lower two eigenvalues are complex and lead to Hopf bifurcations at $n_s \approx 1.567$ and $n_a \approx 1.545$, for symmetric and antisymmetric modes, respectively.  The eigenvalues at each bifurcation are $\sigma = \pm0.912\mathrm i$ and $\sigma = \pm 0.885\mathrm i$ for the symmetric and antisymmetric bifurcations, respectively.}
\label{fig5a}
\vspace{-0.5cm}
\end{figure}

Above a certain value of $n$, there are infinitely many steady states of
(\ref{sseq}), (\ref{bcvdw}). This %fact
was established for ${n=3}$ in \cite{ZL} and recently extended to general $n$
by the authors \cite{DTKFZ}. These solutions are symmetric and can be
arranged (for a given $n$) as a sequence $f_{1}$, $f_{2}$, $\ldots$
according to their values at $\xi =0$ such that $f_{1}(0)>f_{2}(0)>\cdots. $
We can thus depict solution branches as $f_j(0)$ over $n$ (Fig.~\ref{fig1}).
These solutions were computed using the open source numerical continuation
software AUTO-07p\,\cite{AUTO} (see also \cite{TBT,DTKFZ}). As $n$ is
decreased, the solution branches merge, with the first two branches merging
$n=n_{\mathrm c} \approx 1.49915$\,\cite{DTKFZ}.  Our focus is on the change
in dynamics of solutions to (\ref{vdw}) close to this value.

%STABILITY

First, we analyze the stability of $f_1(\xi)$ as $n$ varies.
%This is again carried out using AUTO-07p.
We linearize (\ref{sseq})
about $f_{1}(\xi)$ and seek solutions of the linearized problem in the form
$e^{\sigma \tau }\Phi (\xi )$, obtaining an eigenvalue problem for $%
\sigma = \sigma_R + \mathrm i \sigma_I$.  As $f_1$ is symmetric in $\xi$, $\Phi(\xi)$ may be either symmetric or antisymmetric, which we enforce by applying the appropriate boundary conditions at $\xi=0$, in addition to the far-field conditions arising from (\ref{bcvdw}).  % at a far field value $\xi = L \gg 1$.
For $n=3$, it has been shown \cite{WB12} that there are two trivial modes of perturbation, symmetric with $\sigma_1=1$ and antisymmetric with $\sigma_2=\beta$, which correspond to time and space translation of the singularity, respectively.  Otherwise, all eigenvalues have negative real part, and so $f_1$ is stable.  All other branches $f_2,\, f_3,\, \ldots$ have eigenvalues with positive real part and are unstable. Fig.~\ref{fig5a} displays results for general $n$.  As well as the two trivial eigenvalues, we compute the two nontrivial eigenvalues with largest real part corresponding to symmetric and antisymmetric modes.  These both have negative real part at $n=3$ but increase as $n$ decreases, crossing the imaginary axis at Hopf bifurcations close to $n=n_\mathrm c$ where $f_1$ and $f_2$ merge.  These points are labelled $n_\mathrm s$ and $n_\mathrm a$.  In general, a Hopf bifurcation leads to the existence of a branch of periodic orbits (in scaled time $\tau$, in this case) emanating from the bifurcation.

\begin{figure}
\includegraphics[width=0.45\textwidth]{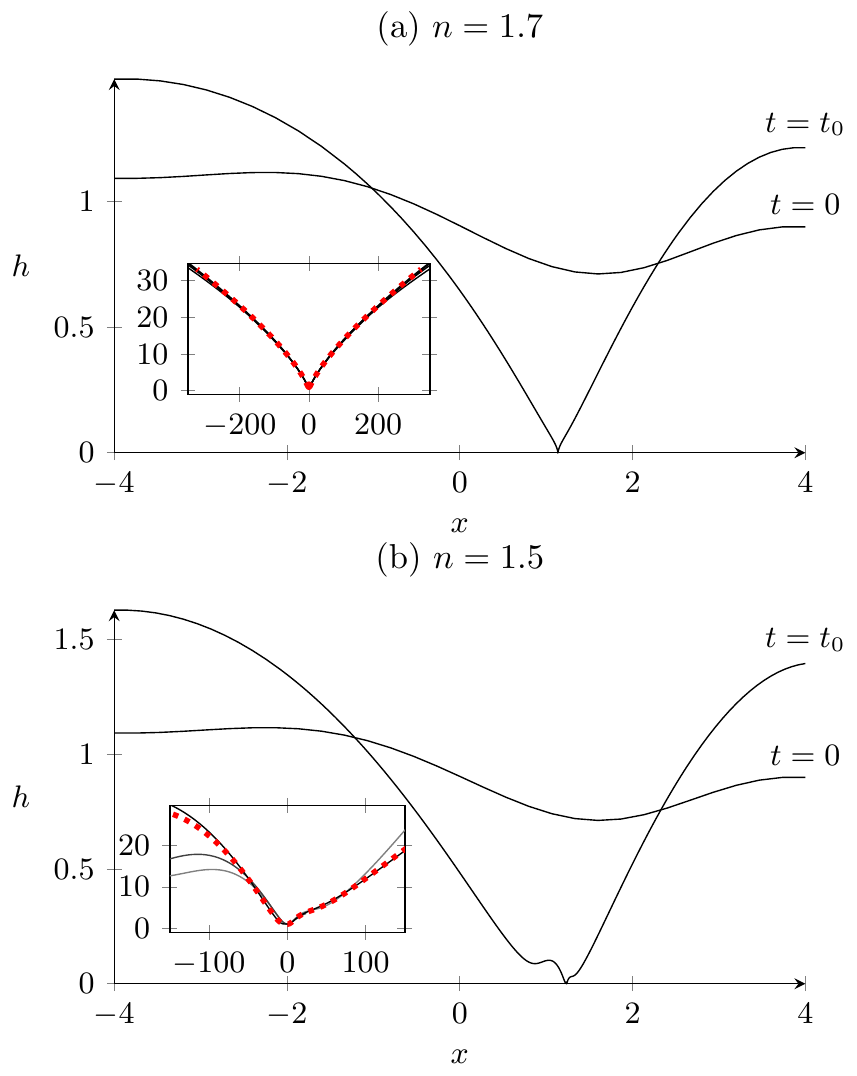}
\vspace{-0.4cm}
\caption{Evolution towards rupture from an initially perturbed profile for (a) $n = 1.7$, and (b) $n=1.5$. Rupture occurs at a point at time $t = t_0$ (note $t_0$ is not the same for each simulation). Inset are interface profiles near the singularity rescaled according to (2), with the dotted lines showing (a)~the stable self-similar solution for $n=1.7$ and (b) the profile on the periodic solution for $n=1.5$ corresponding to the time $\tau$ on the period at which $f_\mathrm{min}(\tau) = \min_\xi f(\xi,\tau)$ is smallest (profiles of the numerical solution are chosen to correspond to this point in the periodic orbit also).  In each case, the scaled behaviour asymptotes to the stable steady state of (\ref{sseq}) for $n=1.7$ and the periodic orbit for $n=1.5$ (see Fig.~\ref{fig4}).}
\label{fig3}
\vspace{-0.5cm}
\end{figure}

%UNSCALED/SCALED NUMERICS
We now explore the implications of this loss of linear stability on the nonlinear dynamics by computation of the time-dependent equation both in the unscaled (\ref{vdw}) and scaled (\ref{sseq}) coordinates. To compute solutions to (\ref{vdw}) that can capture details close to rupture, we implement an adaptive finite difference scheme that increases local mesh refinement near the minimum of $h$ whenever $h_{\min}$ is
less than half of its value at the previous mesh refinement. Fig.~\ref{fig3} shows the results of the computations for (a)~$n = 1.7$ and (b) $n=1.5$, which are on either side of the Hopf bifurcation structure shown in Fig.~\ref{fig1}. Results for other $n$ values are included in the supplementary material. The transition from classical (continuous) self similarity to the onset of cascading oscillations of ge\-o\-met\-ri\-cally decreasing size, is apparent.  The inset in Fig.~\ref{fig3}a shows that the profiles approach a classical self-similar profile [i.e. a steady state of (\ref{sseq})] for $n=1.7$.  In Fig.~\ref{fig3}b, we observe the repetition of the same pattern on geometrically smaller scales, which asymptotically approaches the scaled-time periodic solution to (\ref{sseq}), which we describe next.

The computation of solutions to (\ref{sseq}) is complicated by the trivial eigenvalues corresponding to shifts in space and time.  These instabilities may be thought of as arising from incorrect choices of $x_0$ and $t_0$ in scaling the initial condition.  We remove these instabilities by letting $x_0$ and $t_0$ be time-dependent estimates of the true rupture location, which leads to a new equation of the form
\vspace{-0.2cm}
\begin{equation}
\label{sseq2}
%\hat f_{\hat\xi} = Q(\hat\tau)\left(\alpha\hat f - \beta\hat\xi \hat f_{\hat\xi}\right) + P(\hat\tau)\hat f_{\hat\xi} - \left[ \hat f^{3}\left( \hat f_{\hat \xi \hat \xi}-\frac{1}{n\hat f^{n}}\right) _{\hat \xi }\right] _{\hat \xi },
\hat f_{\hat\xi}\! =\! Q(\hat\tau)\!\left(\alpha\hat f\! -\! \beta\hat\xi \hat f_{\hat\xi}\right)\! +\! P(\hat\tau)\hat f_{\hat\xi} -\! \biggr[\! \hat f^{3}\!\!\left(\! \hat f_{\hat \xi \hat \xi}\!-\!\frac{1}{n\hat f^{n}}\!\right)_{\!\!\hat \xi }\biggl] _{\hat \xi },
\vspace{-0.1cm}
\end{equation}
where $P$ and $Q$ are extra degrees of freedom that may be fixed by applying nonlocal constraints that ensure the rupture remains, or at least asymptotically approaches, $\hat\xi=0$.  We determine $P$ and $Q$ by approximately fixing $\hat f(0,\hat\tau) = 1$ in addition to an integral `pinning' condition \cite{RKML}.  Solutions of (\ref{sseq2}) are scaled back to those of (\ref{sseq}) by
\vspace{-0.2cm}
\begin{equation}
\label{sssub}
%f = Q^\alpha \hat f, \ \xi = Q^{\beta} \hat \xi + \int_0^\tau P(\hat\tau') \, \mathrm d\hat\tau', \ \tau = \int_0^{\hat\tau} Q(\hat\tau') \, \mathrm d\hat\tau'.
f\! =\! Q^\alpha \hat f, \ \xi\! =\! Q^{\beta} \hat \xi\! +\! \int_0^\tau\!\! P(\hat\tau') \, \mathrm d\hat\tau', \ \tau\! =\! \int_0^{\hat\tau} Q(\hat\tau') \, \mathrm d\hat\tau'.
\vspace{-0.1cm}
\end{equation}

\begin{figure}
\includegraphics[width=0.44\textwidth]{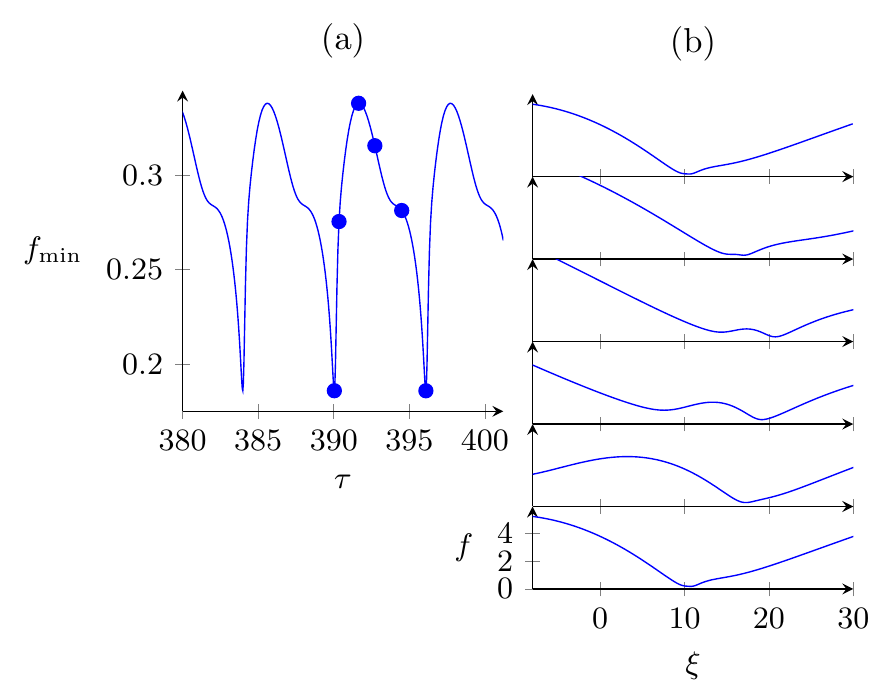}
\vspace{-0.3cm}
\caption{Periodic solutions to (\ref{sseq}), for $n=1.5$:
(a) minimum scaled thickness $f_\mathrm{min}(\tau)$ vs
logarithmic time $\tau$; (b) Solution profiles over one period at points
marked in (a) ($\tau$ increasing from top to bottom).  The profiles show
oscillations with local maxima and minima that advect from the minimum to the
far field.}
\label{fig4}
\vspace{-0.5cm}
\end{figure}

In Fig.~\ref{fig4}, we plot the results of computations of (\ref{sseq2}) for $n=1.5$, transformed to solutions of (\ref{sseq}) via (\ref{sssub}), starting with a generic (asymmetric) initial condition, and run until it is clear that a periodic orbit has developed.  This provides numerical confirmation that stable periodic solutions to (\ref{sseq}) do exist.  We found that development of a periodic orbit is sometimes prevented by rupture occurring away from the origin in (\ref{sseq2}); this is particularly dependent on initial condition and becomes harder to avoid for either $n$ closer to the Hopf bifurcation structure or small values of $n$.  Of particular note in Fig.~\ref{fig4} is that the periodic orbit is asymmetric, and exhibits oscillations that advect outward from the minimum of $f$ to the far field.  These oscillations correspond to the cascade of oscillations, seen in Fig.~\ref{fig3}, that are asymptotically fixed in unscaled space as $t\rightarrow t_0$.

A periodic orbit in self-similar coordinates implies that the rupture of the
film occurs in a discretely, rather than continuously, self-similar
fashion; self-similarity of profiles only holds at discrete times $t_1$,
$t_2,\, \ldots,$ approaching the rupture time $t_0$ geometrically; if $T$
is the period of the orbit, %in self-similar coordinates,
then $t_{N+1}/t_N = e^{-T}$.  Such behavior has been referred to as discrete
self-similarity~\cite{S12} and linked to the existence of periodic orbits in
scaled coordinates; the results in this letter comprise the first explicit
computation of such a periodic orbit~\cite{EF}.  We may understand the
outward propagation of peaks and troughs in the solutions to (\ref{sseq}) in
the scaled coordinates as the creation of `drops' and necks between drops of
geometrically shrinking scale in solutions to the unscaled problem
(\ref{vdw}), thus leading to fractal-like profiles at rupture (as seen in
Fig.~\ref{fig3}b).

The geometric factor in question depends both on $\alpha,\, \beta$ and $T$.  Suppose the maxima $h_{1}$, $h_{2},\,\ldots$
%$h_{N},\,\ldots$
 are located at distances $d_{1}$, $d_{2},\,\ldots$ %$d_{N}$, \ldots\
from $x_{0}$ (with $d_N \rightarrow 0$ as $N\rightarrow\infty$).
Successive maxima %$(d_N,h_N), (d_{N+1},h_{N+1})$
correspond to the same maximum in $(\xi, f)$ at scaled times $\tau$ and $\tau + T$.  Using $h = e^{-\alpha\tau}f$ and ${x-x_0} = e^{-\beta\tau}\xi$, we deduce
${d_{N+1}}/{d_{N}}=e^{-\beta T},\ {h_{N+1}}/{h_{N}}=e^{-\alpha T}.$
%\[
%\frac{d_{N+1}}{d_{N}}=e^{-\beta T},\ \frac{h_{N+1}}{h_{N}}=e^{-\alpha T}.
%\]
The period observed for $n=1.5$ is $T \approx 6.1$, while the periods at the symmetric/asymmetric Hopf bifurcations are $2\pi/0.912 \approx 6.9$ and $2\pi/0.885 \approx 7.1$, respectively.
%{\color{red}\bf (DT: Isn't the asymmetric Hopf bifurcation relevant here, i.e. shouldn't we have ``while the period at the asymmetric Hopf bifurcation is given by $T = 2\pi/(0.885) \approx 7.1$?")}

\begin{figure}
\quad\includegraphics[width=0.44\textwidth]{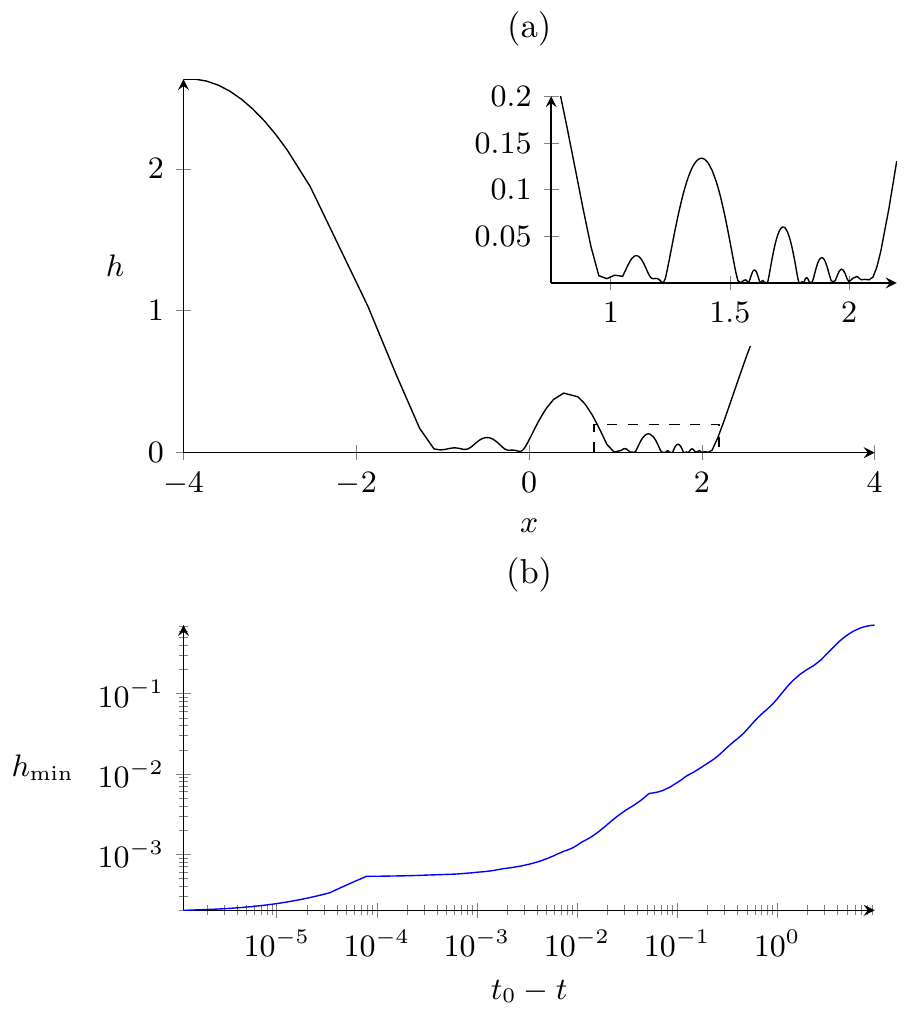}
\vspace{-0.3cm}
\caption{(a) The development of satellite droplet structure for $n=1$, at a late time $t_0$
(inset is the profile where the film is thinnest, showing further subsatellites).
(b) The minimum film thickness as a function of time before  $t_0$;
the observable `kinks' are times at which the position at which minimum thickness is attained changes.}
\label{fig5}
\vspace{-0.5cm}
\end{figure}

Recently the transition from continuous to discrete
and then chaotic self-similar dynamics has been observed in the context of slip
instabilities in elasticity---in particular, simulations of the frictional
sliding dynamics of two elastic bodies in contact\,\cite{V12}, in which there
is a series of Hopf bifurcations on the branch of (self-similar) steady
states. As far as we are aware, Eq.~(\ref{vdw}) is the first model in
hydrodynamics in which a periodic orbit in self-similar variables has been
observed. Our future aim is to systematically compute solution branches from
the Hopf bifurcations.
%While speculative at this point, one possibility is that the system may
%exhibit  a period-doubling path to chaos as $n$ decreases.
While speculative at this point, one possibility is that on such branches there may be
additional bifurcations to quasi-periodic solutions and the system
may become chaotic (via e.g. Ruelle-Takens-Newhouse route)
as $n$ decreases.
For $n=1$ (see Fig.~\ref{fig5}), a cascade of satellite drops of successively
smaller sizes is observed. However, at the space between satellites, new
cascades of subsatellites develop. In between subsatellites at the same
cascade, subsubsatellites develop etc. The result resembles a
fractal-like structure. The minimum height $h_\text{min}$ does not
follow (or oscillate around) a predicatable power law, since the position
where it is reached keeps jumping from one cascade of (sub)satellites
 to another. We do not study this process in detail here, but
present it to demonstrate the complexity that develops as $n$ is
decreased.

As supplemental information we include more details on linear stability and
numerical methods.

\acknowledgments We are grateful to Andrew Archer and
Amparo Galindo for valuable comments and suggestions on disjoining pressure
and intermolecular interactions, respectively. We acknowledge financial
support from the Engineering and Physical Sciences Research Council (EPSRC)
of the UK through Grants No. EP/K041134/1, EP/K008595/1 and EP/L020564/1 and
from the Spanish government through Grant No.~MTM2014-57158-R. MCD was
employed by Imperial College London while undertaking the work reported on in
this paper.


\smallskip

\begin{thebibliography}{99}
\vspace{-0.3cm}

\bibitem{EV} J. Eggers and E. Villermaux,
%Physics of liquid jets,
Rep. Prog. Phys. \textbf{71}, 1 (2008).

\bibitem{General} B. B. Mandelbrot, \emph{The Fractal Geometry of Nature}
    (Macmillan, 1983); A. Perrakis, R. Morris, and V.S. Lamzin, Nat. Struct.
    Mol. Biol. {\bf 6}, 458 (1999); D. Lukkassen, SIAM J. Appl. Math. {\bf
    59}, 1825 (2006).

\bibitem{TSO} M. Tjahjadi, H. A. Stone, and J.M. Ottino,
%Satellite and subsatellite formation in capillary breakup,
J. Fluid Mech.  \textbf{243}, 297 (1992).

\bibitem{SBN12} X. D. Shi, M. P. Brenner, and S. R. Nagel,
%A cascade of structure in a drop falling from a faucet,
Science \textbf{265}, 219 (1994);
M. P. Brenner, X. D. Shi, and S. R. Nagel,
%Iterated instabilities during droplet fission,
Phys. Rev. Lett. \textbf{73}, 3391 (1994) .

\bibitem{CDKOM} H.-C. Chang, E. A. Demekhin, and E. Kalaidin,
%Iterated stretching of viscoelastic jets,
Phys. Fluids \textbf{11}, 1717 (1999);
M. S. N. Oliveira and G. H. McKinley,
%Iterated stretching and multiple breads-on-a-string phenomena in dilute solutions of highly
%extensible flexible polymers.
Phys. Fluids \textbf{17}, 071704 (2005).

\bibitem{Choptuik} M. W. Choptuik, Phys. Rev. Lett.
\textbf{70}, 9 (1993).

\bibitem{Pumir-Siggia} A. Pumir and E. D. Siggia, Phys. Fluids \textbf{4},
    1472 (1992)


\bibitem{CM} R. V. Craster and O. K. Matar,
%Dynamics and stability of thin liquid films,
Rev. Mod. Phys. \textbf{81} 1131 (2009).

\bibitem{ZL} W. W. Zhang and J. R. Lister,
%Similarity solutions for van der Waals rupture of a thin film on a solid substrate,
Phys. Fluids \textbf{11}, 2454 (1999).

\bibitem{WB12}
T. P. Witelski and A. J. Bernoff,
%Stability of self-similar solutions for van der {Waals} driven thin film rupture,
Phys. Fluids, \textbf{11}, 2443 (1999);
T. P. Witelski and A. J. Bernoff,
%Dynamics of three-dimensional thin film rupture,
Physica D \textbf{147}, 155 (2000).

\bibitem{YSKYPKDNGT} P.~Yatsyshin, N.~Savva, and S. Kalliadasis,
%Wetting of prototypical one- and two-dimensional systems: thermodynamics and
%   density functional theory,
J. Chem. Phys. \textbf{142}, 034708 (2015); P.~Yatsyshin, A.O. Parry, and S.
Kalliadasis, J. Phys.: Condens. Matter \textbf{28}, 275001 (2016); S.
Dietrich and M. Napi\'orski,
%Analytic results for wetting transitions in the presence of van der Waals tails,
Phys. Rev. A \textbf{43}, 1861 (1991);
A. Gonz\'alez and M. M. Telo da Gama,
%Density functional theory of long-range critical wetting,
Phys. Rev. E \textbf{62}, 6571 (2000).

\bibitem{BEIMR} D.~Bonn, J.~Eggers, J.~Indekeu, J.~Meunier, and E.~Rolley,
    Rev. Mod. Phys. \textbf{81}, 739 (2009).

\bibitem{Parry_etal} A.O. Parry, J.M.~Romero-Enrique, A. Lazarides, Phys.
    Rev. Lett. {\textbf 93}, 086104 (2004).

\bibitem{DC}
B. V. Derjaguin and N. V. Churaev,
%Structural component of disjoining pressure,
J. Coll. Interf. Sci. \textbf{49}, 249 (1974).

\bibitem{PD} R. M. Pashley,
%Multilayer adsorption of water on silica: An analysis of experimental results,
J. Coll. Interf. Sci. \textbf{78}, 246 (1980); H. T. Davis, \emph{Statistical
Mechanics of Phases, Interfaces and Thin Films} (Wiley, 1995).

\bibitem{HTA} A. P. Hughes, U. Thiele, A. Archer, J. Chem. Phys. \textbf{42},
    074702 (2015).

\bibitem{RAMG}
N. S. Ramrattan, C. Avenda\~no, E. A. M\"uller and A. Galindo,
Molecular Phys. \textbf{113}, 932 (2015).

\bibitem{TDS12} G. F. Teletzke, H. T. Davis, and L. E. Scriven,
%Wetting Hydrodynamics,
Rev. Phys. Appl. \textbf{23}, 989 (1988); G.~F. Teletzke, H.~T.~Davis, and
L.~E. Scriven,
%How liquids spread on solids,
Chem. Eng. Commun. \textbf{55}, 41 (1987).

\bibitem{I} J. N. Israelachvili, \textit{Intermolecular and Surface Forces
    with Applications to Colloidal and Biological Systems} (Academic, New
    York, 1992).

\bibitem{BTARSKRJ}
W.~Boos and A.~Thess,
%Cascade of structures in long-wavelength Marangoni instability}
Phys. Fluids \textbf{11}, 1484 (1999);
A. Alexeev, T. Gambaryan-Roisman, and P. Stephan,
%Marangoni convection and heat transfer in thin liquid films on heated walls with topography: Experiments and numerical study,
Phys. Fluids \textbf{17}, 062106 (2005);
S.~Krishnamoorthy, B.~Ramaswamy, and S.~W.~Joo,
%{Spontaneous rupture of thin liquid films due to thermocapillarity: A full-scale direct numerical experiment},
Phys. Fluids \textbf{26}, 072001 (2014).

\bibitem{YH}
S. G. Yiantsios and B. G. Higgins,
%Rayleigh--Taylor instability in thin viscous films,
Phys. Fluids A \textbf{1}, 1484 (1989).

\bibitem{DTKFZ} M. C. Dallaston, D. Tseluiko, Z. Zheng, M. A. Fontelos, and
    S. Kalliadasis,  Nonlinearity \textbf{30}, 2647 (2017).

\bibitem{AUTO}
E.~J.~Doedel, A.~R.~Champneys, F.~Dercole, T.~F.~Fairgrieve, Y.~A. Kuznetsov, B.~Oldeman, R.~C. Paffenroth, B.~Sandstede, X.~J. Wang, and C.~H.
  Zhang,
\textit{AUTO-07p, Continuation and bifurcation software for ordinary differential equations}, \texttt{http://indy.cs.concordia.ca/auto/} (2007).

  \bibitem{TBT}
{D.~Tseluiko, J.~Baxter, and U.~Thiele},
%A homotopy continuation approach for analysing finite-time singularities in thin liquid films},
IMA J. Appl. Math. \textbf{78}, 762 (2013).

\bibitem{S12}
D. Sornette,
%Discrete-scale invariance and complex dimensions,
Physics Reports \textbf{297}, 239(1998); D. Sornette, \textit{Critical
Phenomena in Natural Sciences: Chaos,Fractals, Self-Organization and
Disorder: Concepts and Tools} (Springer Science \& Business Media 2006).

%\bibitem{K1}
%Y. A. Kuznetsov,
%\textit{Elements of Applied Bifurcation Theory}, \textbf{112}
%(Springer Science \& Business Media 2013).

\bibitem{EF} J. Eggers and M. A. Fontelos, \textit{Singularities: Formation,
    Structure, and Propagation} (Cambridge University Press, 2015).

\bibitem{RKML}
C. W. Rowley, I. G. Kevrekidis, J. E. Marsden and K. Lust, Nonlinearity. \textbf{16}, 1257 (2003).

\bibitem{V12}
R. C. Viesca,
%Self-similar slip instability on interfaces with rate- and state-dependent friction,
Proc. R. Soc. A \textbf{472}, 20160254 (2016);
R. C. Viesca,
%Stable and unstable development of an interfacial sliding instability,
Phys. Rev. E, \textbf{93}, 060202 (2016).


\end{thebibliography}
\end{document}